%% file: main.tex
\documentclass{article}
\usepackage[a4paper,left=2.54cm,right=2.54cm,top=2.54cm,bottom=2.54cm]{geometry}
\usepackage[utf8]{inputenc}
\usepackage{xspace}
\usepackage{amsfonts}
\usepackage{comment} 
\usepackage{amsmath,bm}
\usepackage[nolist]{acronym}
\usepackage{graphicx}
\usepackage{newunicodechar}
\newunicodechar{，}{,}
\usepackage{hyperref}
\usepackage{lineno}
\usepackage{siunitx}
\usepackage{titling}
\usepackage{multirow}
\usepackage{array} 
\usepackage{booktabs} 
\usepackage{rotating}  
\usepackage{nameref}
\usepackage{tabularx} 
\usepackage{caption}  
\input{yuhe_commands.tex}

\title{
An evaluation framework for sparse 4D (3D + time) imaging reconstruction via 
bootstrapped cross-validation
}
\author{Yuhe Zhang$^{1,2,*}$, Zisheng Yao$^1$, Zhe Hu$^1$， Tobias Ritschel$^3$, \\ and Pablo Villanueva-Perez$^1$}
\date{%
    $^1$Synchrotron Radiation Research and NanoLund, Lund University, Box 118, 221 00， Lund, Sweden\\%
    $^2$Medical Radiation Physics, Lund University, Box 117, 221 00, Lund, Sweden\\%
    $^3$University College London, WC1E 6BT London, UK\\[2ex]%
    $^*$yuhe.zhang@sljus.lu.se}

\begin{document}

\maketitle
\begin{abstract}
   Four-dimensional (4D; 3D + time) microscopic imaging has emerged as a powerful technique for investigating dynamic phenomena in complex systems, enabling direct visualization of structural evolution in space and time.
   However, when pushing the limits of spatiotemporal resolution, most time-resolved imaging techniques yield inherently sparse 4D datasets. 
   While deep learning-based reconstruction methods have shown promise in reconstructing 4D from sparse spatiotemporal measurements, a practical approach for evaluating their performance in the absence of a 4D reference has, to the best of our knowledge, been lacking. 
   Here, we present a bootstrapped cross-validation framework that estimates reconstruction performance by quantifying correlations between reconstructions generated from independently sampled subsets of the acquired data, as inspired by the 3D validation strategy in cryo-electron microscopy, where reconstructions from split datasets are compared to assess resolutions.
   This enables both qualitative and quantitative assessment in the absence of ground truth. We investigate two representative scenarios with sparse and ultra-sparse X-ray datasets and validate this approach using 4D-ONIX, a 4D deep-learning reconstruction method, on simulated water droplet collision experiments. 
   The proposed approach provides a reference-free framework for performance estimation and support for better-informed experimental strategies across a wide range of ultrafast imaging applications.
\end{abstract}

\begin{center}
\small \textbf{Keywords:} 4D imaging, reference-free validation, image reconstruction, sparse-view reconstruction, time-resolved microscopy, Deep learning.
\end{center}

\section*{Introduction}

Four-dimensional (4D; 3D + time) microscopic imaging has become an essential tool in science to study the internal structure and evolution of systems that cannot be characterized by surface or two-dimensional (2D) measurements alone. By capturing volumetric snapshots over time, 4D imaging enables the study of dynamic processes, such as deformation, flow, fracture, and phase transitions~\cite{kwon20104d,demene20164d,yashiro2021exploring,goethals2022dynamic}. Advances in experimental imaging technologies, including ultrasound, optical tomography, magnetic resonance imaging (MRI), X-ray tomography, and electron microscopy, have led to an increased demand for high-fidelity 4D reconstruction~\cite{stemkens2018nuts, chin2020magnetic, brown2024four, flannigan20124d, colas2009live}.
Among these imaging technologies, X-ray imaging represents a key area of this broader development. 
Due to its high penetration depth and compatibility with opaque or heterogeneous materials, 4D X-ray imaging enables non-destructive investigation of the internal structure and dynamic behavior of samples opaque to visible light, capturing complex temporal evolution with high spatial resolution.
Recent developments in high-brilliance X-ray sources (e.g., diffraction-limited storage rings~\cite{raimondi2023toward, yao2024new} and X-ray free-electron lasers~\cite{vagovivc2019megahertz}) and high-speed detectors~\cite{mokso2017gigafrost, tsuji2018see}, have opened new opportunities for ultrafast 4D X-ray imaging.
These innovations enable the capture of fast dynamical processes beyond the kilohertz timescale with spatial resolutions down to a few micrometers.
Such imaging capabilities have significantly advanced the study of dynamic phenomena, including fluid flow~\cite{raffel2018particle, kaufer2021stereoscopic, lauriola2019high, mei2019high, villanueva2023megahertz}, material deformation~\cite{garcia2021tomoscopy, buyukozturk2022high, xiang2023ultrasound}, and crack propagation~\cite{wang2024situ}, offering valuable insights into the temporal evolution of complex systems.

Despite these advances, acquiring complete 4D datasets remains a significant challenge due to limitations in sampling rates and acquisition methods, especially when pushing the limits of spatiotemporal resolution.
In scanning-based methods such as time-resolved tomography, projections are acquired sequentially by rotating the sample or the imaging system, making it impractical to obtain a complete angular set at every time point when the object evolves on comparable timescales~\cite{garcia2019using, yashiro2021exploring}. 
Adequate sampling of the object's Fourier space requires angular sampling that satisfies the Nyquist condition~\cite{shannon2006communication}, commonly expressed via the Crowther criterion~\cite{jacobsen2019x}, $N_\theta = \pi D / d$, where $D$ is the maximum size of the imaging object and $d$ is the spatial resolution of the 3D reconstruction. 
For realistic spatial resolutions and object sizes, this corresponds to hundreds to thousands of projections, whereas practical X-ray experiments typically acquire only 10–20\% of this number~\cite{garcia2021tomoscopy,makowska2023operando}. 
Furthermore, high rotational speeds can alter or disrupt the dynamics under observation~\cite{zhang2024advancing}, limiting the applicability of scanning-based approaches.
Scanning-free techniques, including stereo imaging~\cite{hoshino2011development,maire2024dual,shao2025single} and multi-projection or multi-beam configurations~\cite{villanueva2023megahertz,hoshino2013development,villanueva2018hard,voegeli2020multibeam,fainozzi2023three}, address these issues by capturing multiple projections simultaneously without rotating the sample. 
These methods are ideally suited for ultrafast imaging, enabling multi-angle acquisition without scanning the sample~\cite{villanueva2023megahertz,fainozzi2023three}. 
However, practical constraints often limit the number of simultaneous projections to fewer than 5\% of those required by the Crowther criterion, sometimes as few as two or three per time point~\cite{asimakopoulou2023development,rogalinski2025timeresolved3dimagingopportunities}, making the reconstruction problem extremely ill-posed. 
This challenge is not unique to X-rays and arises in many modalities where rapid 4D imaging is required, e.g., electron tomography~\cite{kwon20104d}, MRI~\cite{stemkens2018nuts, chin2020magnetic}, ultrasound~\cite{von1990real,brown2024four}, and optical projection tomography~\cite{rieckher2011microscopic, colas2009live}.

Advances in computational imaging and deep learning have significantly enhanced the capabilities of sparse-view 4D reconstruction. 
Learning-based approaches use differentiable physical operators and neural representations to infer spatiotemporal dynamics from limited or noisy data ~\cite{kimoto2024unsupervised,zhang2024_4donix,hu2025super,friis2025implicit, maas2025nert,zhang2025deep, chawla2025four}.
However, validating such reconstructions remains challenging. 
Acquiring full-reference 4D datasets is often infeasible, and existing non-reference image quality assessment methods either require human scoring or pretrained models learned from large training databases~\cite{liu2017rankiqa,lin2018hallucinated, wang2024crossscore}. 
Therefore, there is an urgent need for reference-free validation strategies that are specifically designed for sparse 4D imaging.

To address this need, we introduce a bootstrapped cross-validation framework for estimating reconstruction fidelity in the absence of ground truth. 
Inspired by statistical bootstrapping~\cite{mooney1993bootstrapping} and 3D validation analyses~\cite{van1987similarity,scheres2012prevention, kalbfleisch2022x}, our approach evaluates the consistency between reconstructions generated from independently sampled subsets of the data. 
This provides an empirical measure of the stability, resolution, and reliability of reconstruction without the need for a reference.
We demonstrate the framework in two representative sparsity regimes that broadly reflect the practical constraints of many imaging modalities:
(i) sparse 4D data capturing 10-20\% of the projections required by the Crowther criterion, which is typical of scanning-based ultrafast tomography, and
(ii) ultra-sparse 4D data capturing less than 5\% of the required views, which is representative of scanning-free configurations.
These two regimes are not exclusive to X-ray imaging, but rather exemplify common experimental trade-offs where the number of projections or experiments is restricted by physical or hardware limitations.
To explain the method, we apply the framework to 4D-ONIX~\cite{zhang2024_4donix}, a 4D deep-learning reconstruction technique capable of recovering 4D dynamics from extremely sparse projections, using simulated water-droplet collision datasets modeled via Navier–Stokes–Cahn–Hilliard equations~\cite{Houseini:17, Lovric2019LowOF}.
Please note that the proposed strategy is not specific to one reconstruction approach and can also be generally applied to other reconstruction approaches using similar bootstrapped cross-validation, such as NeCT~\cite{friis2025implicit}, NeRF-CA~\cite{maas2025nerf}, STRT~\cite{hu2025super}, DYRECT~\cite{goethals2025dyrect}, $X^2$-Gaussian~\cite{yu2025x}, among others. 
4D-ONIX is provided here as an example, and the framework can be applied directly to other 4D reconstruction methods.
By adapting the bootstrapping strategy to sample either projections or experiments, depending on the sparsity regime, and analysing both subset-to-full comparisons and cross-validation from independently sampled 4D inputs, the framework can provide robust qualitative and quantitative estimates of reconstruction fidelity in situations where ground truth is unavailable.

\section*{Method}

In this section, we first outline the assumptions and limitations underlying the proposed 4D validation framework. We then describe the evaluation metrics and the 4D dataset and reconstruction method used for evaluation.
In the end, we present the bootstrapped cross-validation methodology, demonstrated in two representative scenarios: sparse 4D data and ultra-sparse 4D data.

\subsection*{Assumptions on temporal Nyquist sampling}
In most time-resolved imaging, such as in time-resolved X-ray imaging, the achievable temporal resolution is fundamentally constrained by the experimental acquisition system, e.g., the source repetition rate, detector frame rate, or rotational/scanning speed in scanning approaches, which determine the minimum resolvable time interval. 
This temporal sampling interval sets the upper limit on the dynamics that can be faithfully captured and, in turn, influences the attainable spatial resolution, since insufficient temporal sampling leads to averaging of multiple dynamic states and introduces bias in the reconstructions.
In this work, we assume that the temporal domain is sampled at or above the Nyquist rate and focus exclusively on sparsity in the spatial domain. For clarity of analysis, we define the Nyquist velocity, characterizing the maximum resolvable object speed under the assumed temporal sampling conditions.

Although the concept of Nyquist velocity has been defined in some fields---such as Doppler imaging in medical applications---to represent the maximum measurable velocity before aliasing occurs~\cite{nitzpon2002new, sakamoto2016estimation}, an equivalent formulation is largely absent in many imaging fields like ultrafast X-ray imaging. 
According to the Nyquist--Shannon sampling theorem, the maximum detectable frequency of a dynamic process must be less than half of the sampling frequency~\cite{shannon2006communication}. In other words, if the temporal sampling rate is 1 kHz, the highest resolvable dynamic frequency is 500 Hz. 
In 4D imaging, both spatial and temporal sampling shall simultaneously satisfy the Nyquist criterion to faithfully capture dynamic structures. This coupling implies that the maximum measurable velocity is governed by the spatial sampling interval ($\Delta x$) and the temporal sampling step ($\Delta t$), such that 
\begin{equation}
\mathbf{v}_{\text{Nyquist}} = \frac{\Delta x}{\Delta t}. 
\end{equation}
Practically, this requirement implies that the sample’s displacement between two consecutive frames should not exceed one spatial sampling unit (e.g., one pixel), thereby avoiding spatial aliasing and ensuring accurate 4D reconstruction. 
This assumption forms the basis of our study: under a temporally Nyquist-sampled acquisition, we focus on how to quantitatively assess the reliability of 4D reconstructions when spatial sampling is sparse.

Note that some 4D reconstruction algorithms support temporal interpolation~\cite{yao2025physics, RAISSI2019686, gao2025fluidnexus}. 
Interpolation in the temporal domain is usually based on physical priors, such as governing equations, and/or temporal smoothness constraints, such as total variation. 
With this additional prior knowledge, particularly governing physical equations, temporal superresolution becomes achievable. 
However, to obtain a true reconstruction based solely on experimental observation, the effective temporal resolution is fundamentally limited by the Nyquist condition and the actual acquisition rate, regardless of postprocessing. 

\subsection*{4D evaluation metrics}
\label{subsect:metrics}
\begin{table}[hbtp!] 
\centering
\caption{\textbf{Summary of evaluation metrics used for 4D reconstruction assessment.}}
\renewcommand{\arraystretch}{1.35}
\begin{tabularx}{\linewidth}{>{\raggedright\arraybackslash}p{2.5cm} >{\raggedright\arraybackslash}p{1cm} X X}
\toprule
\textbf{Category} & \textbf{Metric} & \textbf{What It Measures} & \textbf{Key Characteristics} \\
\midrule

\textbf{I. Global Error \& Fidelity}
& MSE
& Mean-squared voxel-wise intensity deviation.
& L2-norm-based; penalizes outliers and large errors. \\

& PSNR
& Logarithmic normalization of MSE
& Same as MSE. L2-norm-based; penalizes outliers and large errors.\\

\midrule

\textbf{II. Structural \& Statistical Consistency}
& DSSIM
& Perceptual and structural consistency based on mean, variance, and covariance
& Sensitive to luminance, contrast, and structure; aligns with visual perception. \\

& NMI
& Structural mutual dependence derived from entropy-based functions 
& Robust to non-linear intensity shifts. \\

& NCC
& Structural consistency based on Pearson correlation
& Robust to linear intensity shifts. \\

& FHC
& Frequency-domain structural consistency across 4D hypershells; resolution estimated via the half-bit criterion
& Fourier-domain correlation metric; extends Fourier Shell Correlation to 4D hypershells.  \\

\bottomrule
\end{tabularx}
\label{tbl: metrics}
\end{table}

To quantitatively assess the fidelity of 4D reconstructions, we employ six complementary metrics, as listed in Table \ref{tbl: metrics}.
These metrics can be categorized into two groups based on the aspect of reconstruction quality they measure. First, global error and fidelity are assessed using \ac{MSE} and \ac{PSNR}. \ac{MSE} computes the voxel-wise averaged squared difference between two 4D datasets, providing a straightforward measure of global reconstruction error. \ac{PSNR}, a derived metric based on \ac{MSE}, evaluates the signal's peak intensity relative to the noise level, offering a quantitative measure of image quality. Second, structural and statistical consistency are evaluated using \ac{DSSIM},  \ac{NMI}, \ac{NCC}, and \ac{FHC}. \ac{DSSIM}, the dissimilarity form of the structural similarity index, evaluates perception-based differences in real space, capturing contrast and structural discrepancies that are not well represented by simple intensity error. \ac{NMI} measures the shared information between two reconstructions, assessing the statistical dependence of the reconstructed dynamics. 
\ac{NCC} is a statistical measure that quantifies the linear similarity or agreement between two reconstructed 4D volumes.
Finally, frequency-domain resolution is determined by \ac{FHC}. 
It generalizes the Fourier shell correlation (FSC)~\cite{VANHEEL2005250} from 3D to 4D by calculating frequency-domain correlations across 4D hypershells in reciprocal space. 
As the temporal dimension is not directly comparable to spatial dimensions, all frequency radii are normalized such that the maximum spatial–temporal frequency corresponds to the Nyquist limit.
The half-bit criterion~\cite{VANHEEL2005250} is then applied to FHC curves to estimate resolution, ensuring a statistically meaningful threshold for interpretability. 
Each metric is computed directly in 4D, avoiding reductions to 3D or 2D, and thereby providing a faithful characterization of spatiotemporal reconstruction accuracy. 
 
\subsection*{4D dataset and reconstruction method for evaluation}
To assess the proposed validation framework, we employed a simulated 4D dataset with known ground truth. 
Specifically, we used a water droplet collision dataset~\cite{4D-ONIX-data} comprising 16 collision experiments with 10\% variation in droplet size and impact velocity.
The simulations were generated using the Navier--Stokes--Cahn--Hilliard formulation implemented in the open-source DUNE framework, in particular DUNE-FEM~\cite{dunereview:21, dunefemdg:21}.  
Each experiment consisted of 75 time steps of $128^3$\, voxel 3D volumes, sampled according to the temporal Nyquist velocity requirement, capturing the full sequence of two droplets approaching, colliding, and coalescing.
We emulated data from a time-resolved X-ray tomography experiment.
For that, we generated projections from 16 equally spaced angles within \ang{180} ( from \ang{0} to \ang{168.75}) at each time point for every experiment, which is approximately 10\% of the projections required by the Crowther criterion. 

For reconstruction, we employed 4D-ONIX~\cite{zhang2024_4donix}, a state-of-the-art 4D X-ray imaging method based on neural implicit representations~\cite{Mildenhall2020NERF}. 
The model uses multiple neural networks to extract features from all input projections (with known geometry) and jointly optimizes over all frames and experiments. 
After training, it produces time-resolved 3D volumes from a sparse set of projections at each time point.
It should be noted that 4D-ONIX is used here merely as an example reconstruction framework for sparse and ultra-sparse 4D data.
The proposed validation methodology is general and not specific to any reconstruction approach. 

To evaluate generalization and enable independent validation, we simulated an additional droplet collision experiment using the same physical model. 
This test case lies within the distribution of the training data, exhibiting a 5\% variation in droplet size and velocity. 
As in training, 16 projections uniformly spanning \ang{180} were generated for this unseen experiment.
For the droplet collision dynamics considered here, the available dataset was sufficient to capture the essential dynamics. 
However, more datasets may be required for reconstructing higher-fidelity dynamics or more complex physical processes.

\subsection*{Performance estimation via bootstrapped cross-validation}

In this section, we estimate the performance using the proposed bootstrapped cross-validation method.
Inspired by bootstrapping and tomographic resolution assessment techniques, where projection images are split into two independent subsets for resolution estimation~\cite{scheres2012prevention}, we extended this idea to 4D reconstructions. 

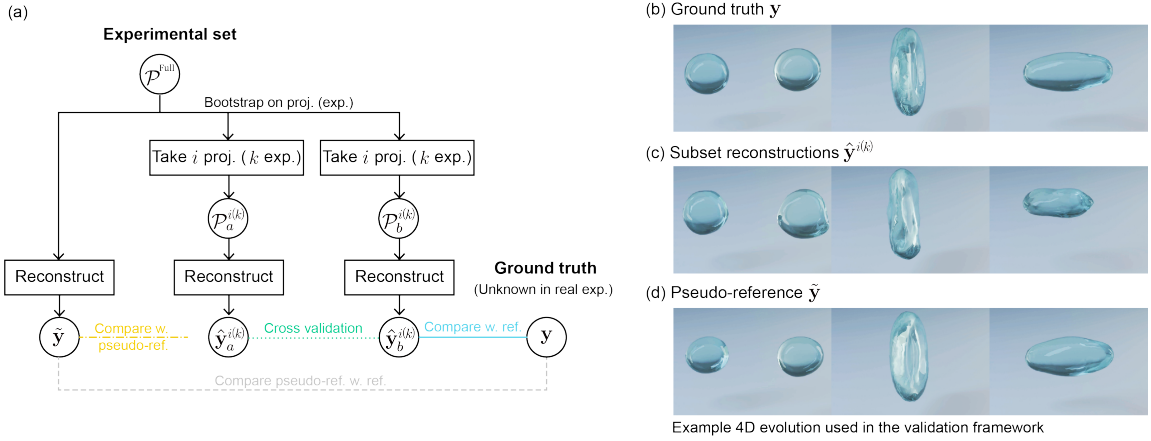
\myfigure{flow_chart}{Schematic of the bootstrapped cross-validation framework. (a) Illustration of the workflow for the proposed framework applying to sparse (or ultra-sparse) 4D reconstruction. Starting from an experimental dataset consisting of 16 experiments with 16 projections per time point, bootstrapping is performed by selecting $i$ projections (or $k$ experiments) from the full set. A reconstruction from all 16 projections yields a full-set reconstruction \fullsetRecon, which serves as a pseudo-reference when ground truth \groundTruth is unavailable in real experiments. Independently bootstrapped subsets $\projSet^{i(k)}$ are used to train multiple models, producing a collection of subset reconstructions $\hat{\mathbf{y}}^{\,i(k)}$. Two examples, $\hat{\mathbf{y}}_a^{\,i(k)}$ and $\hat{\mathbf{y}}_b^{\,i(k)}$, are shown in the figure. Performance is evaluated in three ways: (i) comparison of subset reconstructions $\hat{\mathbf{y}}^{i(k)}$ to ground truth $y$, (ii) comparison of $\hat{\mathbf{y}}^i$ to the pseudo-reference \fullsetRecon, and (iii) cross-validation by computing similarity metrics between a pair of independently reconstructed 4D volumes ($\hat{\mathbf{y}}_a^{\,i(k)}$, $\hat{\mathbf{y}}_{b}^{\,i(k)}$).
(b–d) Example 4D evolution of the water droplet collision dataset used in the validation framework. The images show early, intermediate, and late stages of the dynamic process for (b) the ground truth \groundTruth, (c) the subset reconstructions $\hat{\mathbf{y}}^{i(k)}$, and (d) the pseudo-reference reconstructed from the full available experimental dataset \fullsetRecon. 
Here, the subset reconstruction is an example of a reconstruction using a single experiment with four projections, and the pseudo-reference is based on 16 experiments with four projections, as demonstrated in the ultra-sparse 4D data case.
}

Let $ \groundTruth \in \mathbb{R}^{T \times X \times Y \times Z}$ denote the true 4D object, and $\mathcal{D}_n$ denote a dataset containing $n$ independent low-dimensional measurements of \groundTruth, $\forwardModel_{\boldsymbol{\theta}}$ denote a reconstruction model with parameters
$\boldsymbol{\theta}$. 
Training on $\mathcal{D}_n$ yields an estimator
$\hat{\boldsymbol{\theta}}_n$ and a reconstructed 4D volume
\begin{equation}
\hat{\mathbf{y}}_n = \forwardModel_{\hat{\boldsymbol{\theta}}_n}(\mathcal{D}_n).
\end{equation}
Under the assumption that samples in $\mathcal{D}_n$ are independent in both space and time, the estimated parameters  $\hat{\theta}_n$  of the reconstruction model converge in probability to the true parameter values $\theta$, as the number of data points $n$ increases, i.e.~\cite{goodfellow2016deep},

\begin{equation}
\operatorname{plim}_{n \to \infty} \hat{\boldsymbol{\theta}}_n = \boldsymbol{\theta}.
\label{eq:plim}
\end{equation}
Consequently, reconstructions obtained from sufficiently large and independent datasets converge to the same underlying 4D solution:
$\operatorname{plim}_{n \to \infty} \hat{\mathbf{y}}_n = \groundTruth$.
Therefore, in practical scenarios when the reference is unavailable, a deterministic reconstruction obtained from the full available dataset can serve as a pseudo-reference, given that the observation points $n$ are sufficient for the specific 4D distribution. We denote the pseudo-reference as \fullsetRecon.

The workflow of the bootstrapped cross-validation framework for sparse 4D reconstruction and example frames of the 4D data are illustrated in Figure \ref{fig:flow_chart}.
To enable performance estimation without a reference, we further introduce a randomized sampling scheme. 
Let 
$\projSet^{i(k)} \sim \mathcal{D}$
denote a random subset of measurements drawn from the probability distribution $\mathcal{D}$. Depending on the sparsity regime, this may correspond to random selection of $i$ projection angles or random selection of $k$ experiments, as shown in Figure \ref{fig:flow_chart}.
The corresponding reconstruction $\hat{\mathbf{y}}^{i(k)} = \forwardModel^{i(k)}[\projSet^{i(k)}]$ is therefore a random estimator of the true 4D object \groundTruth. Consider a pairwise cross-validation estimator, with $\hat{\mathbf{y}}_a^{\,i(k)}$ and $\hat{\mathbf{y}}_{b}^{\,i(k)}$ be two independent reconstructions from the two independent sampled subsets $\projSet^{i(k)}_a$ and $\projSet^{i(k)}_b$ . 
The cross-validation metric is defined as 
\begin{equation}
    C = M(\hat{\mathbf{y}}_a^{\,i(k)}, \hat{\mathbf{y}}_b^{\,i(k)}),
    \label{Eq: correlation}
\end{equation}
where $M(\,\cdot,\cdot)$ denotes a 4D evaluation metric, e.g., MSE, DSSIM, etc.
The expectation
$\mathbb{E}_{\mathcal{P}_a, \mathcal{P}_b \sim \mathcal{D}} [M(\hat{\mathbf{y}}_a^{\,i(k)}, \allowbreak \hat{\mathbf{y}}_b^{\,i(k)})]$
and variance $\text{Var} _{\mathcal{P}_a, \mathcal{P}_b \sim \mathcal{D}}[M(\hat{\mathbf{y}}_a^{\,i(k)}, \allowbreak \hat{\mathbf{y}}_b^{\,i(k)})]$ provide
estimations of the performance and reproducibility of the reconstruction process under independent data realizations.

Here, we first considered a scenario for sparse 4D data, where approximately 10\% of projections required by the Crowther criterion were captured, as is common for scanning-based 4D X-ray imaging approaches such as time-resolved X-ray tomography.
Specifically, we randomly sampled $i \in \{2,4,8\}$ projections from the full 16-angle set and used these subsets for model optimization, ensuring the selected projections for each subset were evenly spaced within \ang{180}, as $\projSet^i = \left\{ \ang{0}, \frac{\ang{180}}{i},..., \ang{180}-\frac{\ang{180}}{i} \right\}$.
For instance, $i=2$ corresponds to \ang{0} and \ang{90}.
The single-projection case was excluded because the reconstruction becomes dominated by the input view, lacking any shared 3D information.
For each sampling size \(i\), 100 independent subsets were generated,  each containing 16 experiments with $i$ projections. 
Applying a trained model $\forwardModel^i$ to any projection set \( \projSet^i \) yielded a 4D reconstruction $\hat{\mathbf{y}}^i=\forwardModel^{i}(\projSet^{i})$.

The resulting reconstructions were evaluated in two ways. 
First, each subset reconstruction $\hat{\mathbf{y}}^{i}$ was compared against the ground truth \groundTruth\ and against the reconstruction obtained from all 16 projections, which is used as the pseudo-reference \fullsetRecon. 
These results are shown in Figure \ref{fig:IT_CV_Sparse}, where blue solid lines correspond to the subset-to-ground-truth comparison and yellow dashdotted lines correspond to the subset-to-fullset comparison. 
The gray dashed lines indicate the performance of the 16-projection fullset reconstruction relative to the ground truth.
Second, we evaluated the cross-validation of $\hat{\mathbf{y}}^i$ pairs reconstructed using independent 4D data.
Specifically, 1000 pairs of subset reconstructions $\hat{\mathbf{y}}^{i}$ were randomly selected and compared. 
To ensure temporal independence between the paired inputs, we employed interlaced time points, i.e., in addition to Equation \ref{Eq: correlation}, the correlation metric is evaluated as

\begin{equation}
C = M(\hat{\mathbf{y}}_a^i|_{t=2N},\,\hat{\mathbf{y}}_b^i|_{t=2N+1}),
\label{Eq:interlaced}
\end{equation}
where $N \in \mathbb{N}$ stands for non-negative integers.
Given that the dataset contains 75 frames, the last frame was skipped to maintain consistency. 
The resulting cross-validation statistics are shown as green dotted lines in Figure \ref{fig:IT_CV_Sparse}. 
For completeness, a comparison using non-interlaced time points is provided in the Supplementary Section S1.
As summarized in the section \nameref{subsect:metrics}, six evaluation metrics were used. 
Mean values are indicated by markers, with error bars showing standard deviations. MSE and DSSIM are plotted on a logarithmic scale.

\mycfigure{IT_CV_Sparse}{
Performance evaluation of 4D reconstructions under sparse spatial sampling. 
The performance of the reconstructions is shown as a function of the number of projections.
The six metrics used are MSE, DSSIM, FHC, PSNR, NCC, and NMI, with error bars indicating standard deviations across 100 subset reconstructions and 1000 cross-validation pairs.
Three comparisons are shown: subset reconstructions $\hat{\mathbf{y}}^i$ versus ground truth \groundTruth (circles, blue solid lines), subset reconstructions $\hat{\mathbf{y}}^i$ versus the full 16-projection reconstruction \fullsetRecon (triangles, yellow dashdotted lines), and cross-validation between independently reconstructed 4D volumes (squares, green dotted lines). 
The gray dashed line denotes the baseline performance of the full 16-projection reconstruction \fullsetRecon relative to the ground truth \groundTruth. 
}

Second, we considered a scenario for ultra-sparse 4D data.
In many practical scenarios, acquiring even 10--20\% of the projections required by the Crowther criterion at each time point is infeasible, particularly in scanning-free 4D imaging systems. 
Such setups typically provide fewer than 5\% of the required projections, often offering fewer than five simultaneous views.
To assess reconstruction performance under these conditions, we performed a bootstrapped cross-validation study based on varying the number of experiments used for model optimization.
An experiment refers to an independent realization of a similar or quasi-reproducible dynamic process. 
Each experiment is acquired with the same number of ultra-sparse projections. 

To emulate scanning-free acquisition, we generated four projections spaced \ang{45} apart for each experiment by subsampling the original 16-angle dataset. 
Because scanning-free systems may not preserve the exact sample orientation, and practical implementations often allow mechanical variability (e.g., uncalibrated sample rotation), each experiment was assigned a different angular offset. 
This sampling strategy mimicked the type of data produced by multi-projection configurations such as those in~\cite{asimakopoulou2023development,maire2024dual,liang2023sub}.
Across the 16 experiments, these resulted in a dataset representative of an imaging system capable of capturing four simultaneous projections whose absolute orientations were unknown. 
A complete list of the projection angles assigned to all 16 experiments is given in Supplementary Table S1.
Importantly, the neural network was never provided with the projection angles; instead, it only receives the four intensity images per experiment.

To study the impact of dataset size, we randomly sampled \(k \in \{1, 2, 4, 8\}\) experiments from the full set of 16. 
For each \(k\), we generated 100 independent subsets and trained a separate model for each subset. Applying a trained model \(\forwardModel^{(k)}\) to the validation experiment yielded a 4D reconstruction $\hat{\mathbf{y}}^{(k)} = \forwardModel^{(k)}(\projSet^{(k)})$,
where \(\projSet^{(k)}\) denotes the four-projection measurements from the selected $k$ experiments.

The results are presented in Figure~\ref{fig:IT_CV_UltraSparse}. 
Similarly, the resulting reconstructions were evaluated against both the full 16-experiment reconstruction (\fullsetRecon) and the ground truth (\groundTruth) using the six metrics described previously. Cross-validation statistics were also computed from pairs of independently reconstructed 4D datasets. 
The gray dashed lines indicate the performance of the full 16-experiment reconstruction relative to the ground truth.

\mycfigure{IT_CV_UltraSparse}{
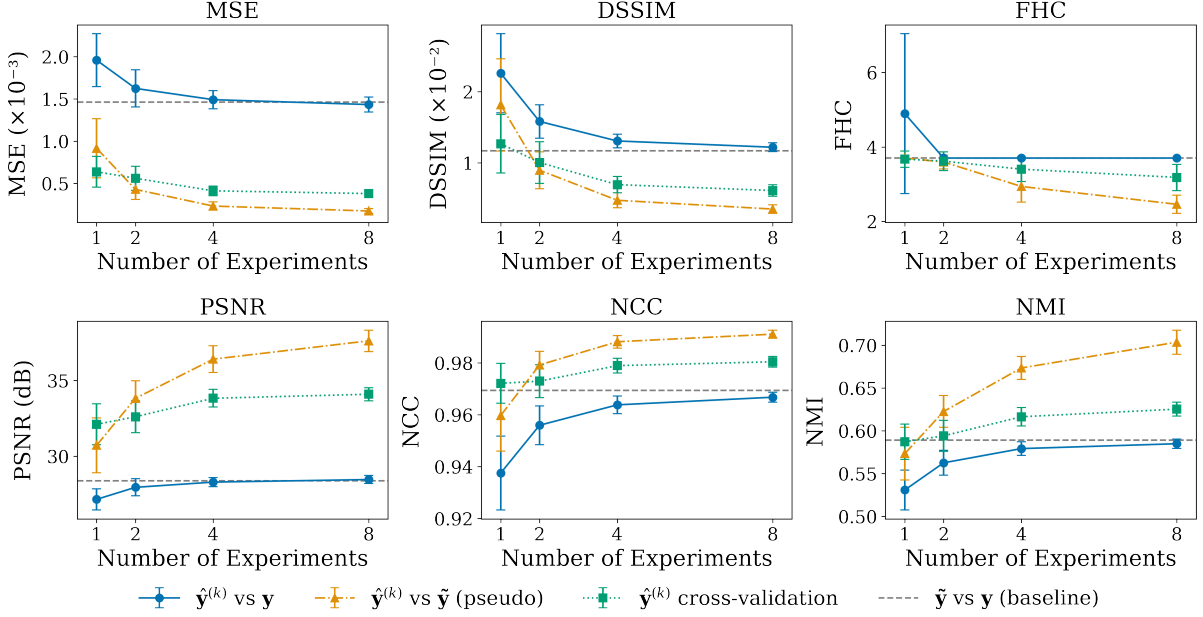
Performance evaluation of 4D reconstructions under ultra-sparse spatial sampling. 
The performance of the reconstructions is shown as a function of the number of experiments.
The six metrics used are MSE, DSSIM, FHC, PSNR, NCC, and NMI, with error bars indicating standard deviations across 100 subset reconstructions and 1000 cross-validation pairs. 
Three comparisons are reported: subset reconstructions $\hat{\mathbf{y}}^i$ versus ground truth \groundTruth (circles, blue solid lines), subset reconstructions $\hat{\mathbf{y}}^i$ versus the full 16-experiment reconstruction \fullsetRecon (triangles, yellow dashdotted lines), and cross-validation between independently reconstructed 4D volumes (squares, green dotted lines). 
The gray dashed line denotes the baseline performance of the full 16-experiment reconstruction \fullsetRecon relative to the ground truth  \groundTruth. 
}

\section*{Results and Discussion}

In this study, we investigated a bootstrapped cross-validation framework for estimating the performance of sparse 4D reconstructions in scenarios where full 4D ground truth is unavailable or difficult to obtain. Two regimes were considered: (i) sparse 4D data, where each time point contained approximately 10–20\% of the projections required by the Crowther criterion (typical for scanning-based imaging), and (ii) ultra-sparse 4D data, where fewer than 5\% of projections were available, as in scanning-free systems.

The schematic of the bootstrapped cross-validation framework and example reconstructions are shown in Figure \ref{fig:flow_chart}.
For the sparse-view scenario, bootstrapping was performed across \(i \in \{ 2, 4, 8\}\) projections. 
For each projection count $i$, 100 independently sampled subsets were used to train 100 separate models, producing 100 independent reconstructions. 
These reconstructions were evaluated against the ground truth \groundTruth and against the 16-projection reconstruction \fullsetRecon, which serves as a pseudo-reference when a true reference is unavailable. 
As shown in Figure \ref{fig:IT_CV_Sparse}, reconstruction performance improved consistently as the projection count increased, and the variance across the 100 reconstructions decreased accordingly. 
For only two projections, the problem remained strongly under-determined, and independently trained models converged to similarly biased solutions despite being far from the ground truth. 
When eight projections were used, all metrics approached the 16-projection pseudo-reference, and the standard deviations fell below 1\%.
In addition, 1000 pairs of subset reconstructions $\hat{\mathbf{y}}^i$ were randomly selected for cross-validation. This provided a reliable quantitative estimate of reconstruction quality once eight projections were used, where both the mean performance and variance became stable. 
Comparing Figure~\ref{fig:IT_CV_Sparse} with Supplementary Figure S1, we noticed that the agreement computed using interlaced time points was consistently lower than that obtained from non-interlaced cross-validation.
This is expected, as interlaced comparisons inherently introduce temporal offsets between reconstructions, capturing both reconstruction consistency and physical variation in the dynamics.
This highlights that valid cross-validation requires independent 4D inputs, not merely independent spatial configurations.
For the ultra-sparse scenario, where each time point contained only four projections, we performed bootstrapping across the number of experiments used for training. 
As shown in Figure \ref{fig:IT_CV_UltraSparse}, reconstruction quality improved markedly as the number of experiments increased, and the variance across the 100 reconstructions decreased in parallel.
Four experiments provided stable performance across all metrics. 
With eight experiments, the results closely approached the 16-experiment baseline.

Across both regimes, reconstruction performance can be assessed using several complementary strategies:
(i) Subset-to-full comparison: evaluating subset reconstructions against the corresponding full-set reconstruction.
As stated in Equation~\ref{eq:plim}, the model parameters estimated from independent 4D data converge in probability as the dataset size increases, provided that the samples are independent in both space and time.
If the performance curves approach a plateau with the increasing number of projections or experiments, the reconstruction is close to optimal and can serve as a pseudo-reference for performance evaluation. 
Otherwise, additional input data is needed for achieving more reliable results.
(ii) Interlaced-time cross-validation: computing correlations between reconstructions generated from 4D inputs sampled independently. 
This requires true 4D independence, i.e., independence across both space and time, to avoid overly optimistic estimates.
(iii) Variance convergence: monitoring the standard deviation across the 100 reconstructions. For all metrics, the variance decreases monotonically and falls below 1\% once a plausible and stable reconstruction regime is reached.
In all approaches, increasing the amount of independent 4D data systematically improves the robustness and reliability of the quantitative performance estimation.   

Within this validation framework, structural and statistical metrics are generally more reliable than global-error fidelity metrics (MSE, PSNR), especially in the ultra-sparse regime. Global-error and fidelity metrics are dominated by local intensity deviations and are sensitive to global scaling or bias, whereas correlation metrics capture spatial–temporal coherence and frequency-domain criteria quantify resolvable detail directly in Fourier space. We recommend using at least two complementary metrics for practical reference: NCC in real space and FHC in frequency space. For FHC, we use the highest frequency present in 4D or 3D space as the Nyquist frequency.
Compared to conventional Fourier correlation-based analysis, this choice can produce more conservative results. Depending on the experimental and sampling conditions, using a less stringent definition of the Nyquist frequency may provide more practical and meaningful results.

A characteristic feature noticeable in both Figure \ref{fig:IT_CV_Sparse} and Figure \ref{fig:IT_CV_UltraSparse} is the intersection between the $\hat{\mathbf{y}}$ vs \groundTruth curves (yellow dashdotted) and the $\hat{\mathbf{y}}$ cross-validation curves (green dotted). 
When only two projections or experiments are available, the reconstruction problem is under-determined, so independently trained models tend to generate reconstruction artifacts that are highly correlated but different from the ground truth.
As more projections or experiments are included, the subset reconstructions improve rapidly, increasing their agreement with the pseudo-reference, while cross-validation becomes a stricter measure because it compares reconstructions generated from two fully independent 4D inputs rather than a fixed reference. 
These opposing trends produce intersection points between the $\hat{\mathbf{y}}$ vs \groundTruth (yellow dashdotted) and $\hat{\mathbf{y}}$ cross-validation (green dotted) curves, and such intersections indicate that the reconstructions are beginning to stabilize. 
When the projection or experiments count approaches the full set (e.g., eight projections), the $\hat{\mathbf{y}}$ vs \groundTruth curves (yellow dashdotted) outperform the pseudo-reference baseline (gray dashed) in both scenarios for all metrics， this is because both reconstructions share similar model-induced biases and reconstruction priors, leading to overfitting and over optimistic estimations.
In contrast, cross-validation remains more conservative by using two independently trained models.
Although not reported here, correlating extremely low-data reconstructions (e.g., two-projection or two-experiment cases) against reconstructions obtained with higher projection or experiment counts could further disentangle correlations arising from shared reconstruction artifacts, providing a complementary validation.

There are, however, several limitations in the present study. First, the validation framework has only been demonstrated on simulated water droplet collision data and has not been tested on real experimental datasets, which may exhibit higher noise levels and other reconstruction artifacts. The noisy nature of experimental data may introduce additional uncertainty into the evaluation process. Nevertheless, the bootstrapping strategy can be extended to datasets from diverse experimental contexts, including material deformation, crack propagation, and cavitation dynamics.
Secondly, although the physics governing water droplet collision dynamics is known a priori, this information was not incorporated into the validation process. 
The reconstructions do not include key physical properties required by the Navier–Stokes equations, such as the velocity and pressure fields, which limits the direct implementation of physics-based validation~\cite{yao2025physics}.
For future work, physics-based constraints such as mass conservation, energy conservation, or dynamical models can be explored as complementary validation strategies~\cite{RAISSI2019686}.
Finally, while the proposed framework allows for the quantitative estimation of reconstruction performance without a reference, the accuracy of these estimates is inherently dataset-dependent and challenging to determine precisely. 
In general, larger datasets enhance estimation reliability, so incorporating as many independent measurements as possible is essential for the proposed validation framework.

With the continuous advancement of experimental hardware and computational tools for time-resolved imaging, sparse-view or even ultra-sparse-view reconstructions without an available ground truth will become increasingly common~\cite{kwon20104d,demene20164d,yashiro2021exploring, mokso2017gigafrost, goethals2025dyrect, gao2024dynamic, rosen2024synchrotron,maire2024dual}. 
Evaluating reconstruction quality in the absence of ground truth is inherently challenging. 
In this work, we introduced a bootstrapped cross-validation framework for deep-learning-based ultrafast 4D X-ray imaging reconstruction. This approach enables reliable qualitative and quantitative evaluation of 4D reconstruction methods in practical scenarios where direct ground truth measurements are inaccessible. We demonstrated, for both sparse and ultra-sparse 4D data, that performance can be approximated by measuring the correlation or similarity between pairs of reconstructions obtained from independently sampled training or validation datasets. 
Depending on experimental constraints, bootstrap studies can be conducted on the number of projections or the number of experiments available.
Although not explored here, sparse-view studies may also incorporate angular bootstrapping by applying a trained model to an alternative validation set with larger angular offsets to probe generalization under rotation shifts. 
Our results show that bootstrapped cross-validation provides meaningful performance estimates even without ground truth, offering a practical tool for assessing and guiding experimental design. 
This framework sets the groundwork for more reliable reconstructions and better-informed experimental strategies across a broad range of ultrafast imaging applications.

\section*{Acknowledgement}
We are grateful to Z. Matej for his support and access to the GPU‐computing cluster at MAX IV. 
We are also grateful to Robert Klöfkorn for simulating the water droplet collision datasets.
This work has received funding by ERC‐2020‐STG 3DX‐FLASH 948426.

\bibliographystyle{unsrt}
\bibliography{main}

\newpage

\appendix
\counterwithin{figure}{section}

\input{supp}



\end{document}

%% file: yuhe_commands.tex
\newcommand{\mymath}[2]{\newcommand{#1}{\TextOrMath{$#2$\xspace}{#2}}}
\mymath{\frcLossWeight}{\lambda_\mathrm{FRC}}
\mymath{\ssimLossWeight}{\lambda_\mathrm{SSIM}}
\mymath{\gpLossWeight}{\lambda_\mathrm{GP}}

\mymath{\encoder}{\mathrm{\mathbf E}}
\mymath{\mlp}{\mathrm{\mathbf F}}
\mymath{\discriminator}{\mathrm{\mathbf D}}
\mymath{\generator}{\mathrm{\mathbf G}}
\mymath{\dataDistribution}{p_D}
\mymath{\randomDistribution}{p_\nu}
\mymath{\latentVector}{\mathbf{z}}
\mymath{\firstAngle}{\alpha}
\mymath{\secondAngle}{\beta}
\mymath{\viewAngle}{\nu}
\mymath{\realAngle}{v}
\mymath{\predictContrast}{\mathbf{\hat{c}_\nu}}
\mymath{\realContrast}{\mathbf{c}_v}
\mymath{\contrast}{\mathbf{c}}
\mymath{\spatialCoordinates}{\textit{\textbf{x}}}
\mymath{\temporalCoordinates}{\textit{t}}
\mymath{\refractiveIndex}{n}

\mymath{\fullsetRecon}{\tilde{\mathbf{y}}}
\mymath{\groundTruth}{\mathbf{y}}
\mymath{\reconAfix}{\hat{\mathbf{y}}_1^i}
\mymath{\reconBfix}{\hat{\mathbf{y}}_2^i}
\mymath{\reconAshift}{\hat{\mathbf{y}}_1^{i\,\prime}}
\mymath{\reconAdiffExp}{\hat{\mathbf{y}}_1^{(k)}}
\mymath{\reconBdiffExp}{\hat{\mathbf{y}}_2^{(k)}}
\mymath{\fullsetRecondiffExp}{\hat{\mathbf{y}'}}
\mymath{\forwardModel}{\boldsymbol{\mathcal{F}}}
\mymath{\projSet}{\mathcal{P}}
\mymath{\expSet}{\mathbf{E}}

\begin{acronym}
\acro{DL}{Deep Learning}
\acro{IoR}{Index of Refraction}
\acro{ONIX}{Optimized Neural Implicit X-ray imaging}
\acro{NeRF}{Neural Radiance Fields}
\acro{CNN}{Convolutional Neural Network}
\acro{MLP}{Multilayer Perceptron}
\acro{MSE}{Mean Squared Error}
\acro{GAN}{Generative Adversarial Network}
\acro{XMPI}{X-ray Multi-Projection Imaging}
\acro{SASE}{Self-Amplified Spontaneous Emission}
\acro{ESRF}{European Synchrotron Radiation Facility}
\acro{SR}{Synchrotron Radiation}
\acro{MHz-XMPI}{MHz X-ray Multi-Projection Imaging}
\acro{PSF}{Point Spread Function}
\acro{MAE}{Mean Absolute Error}
\acro{ML}{Machine Learning}
\acro{FRC}{Fourier Ring Correlation}
\acro{FSC}{Fourier Shell Correlation}
\acro{FHC}{Fourier Hypershell Correlation}
\acro{SSIM}{Structural Similarity Index Measure}
\acro{PSNR}{Peak Signal-to-Noise Ratio}
\acro{NMI}{Normalized Mutual Information}
\acro{NCC}{Normalized Cross-Correlation }
\acro{3D}{Three Dimensional}
\acro{XFEL}{X-ray Free-Electron Laser}
\acro{European XFEL}{European X-Ray Free-Electron Laser Facility}
\acro{SPring-8}{Super Photon Ring – 8 GeV}
\acro{ONIX}{Optimized Neural Implicit X-ray imaging}
\acro{CNN}{Convolutional Neural Network}
\acro{MLP}{Multilayer Perceptron}
\acro{GAN}{Generative Adversarial Network}
\acro{DSSIM}{Dissimilarity Structure Similarity Index Metric}
\acro{PDE}{Partial Differential Equation}
\acro{CDI}{Coherent Diffraction Imaging}
\acro{PINN}{Physics-Informed Neural Network}
\acro{IoR}{Index of Refraction}
\end{acronym}

\def\figurePath{Figures/}

\def\figurePath{Figures/}

\def\myfigure#1#2{%
    \begin{figure}[htbp!]%
    \centering\includegraphics*[width = 0.95\linewidth]{\figurePath#1}%
    \vspace{-.2cm}%
    \caption{#2}%
    \vspace{-.3cm}%
    \label{fig:#1}%
    \end{figure}%
}

\def\mycfigure#1#2{%
    \begin{figure*}[htbp!]%
    \centering\includegraphics*[width = \linewidth]{\figurePath#1}%
    \vspace{-.2cm}%
    \caption{#2}%
    \label{fig:#1}%
    \end{figure*}%
}

\newcommand{\change}[2]{%
\hypertarget{hyperline:#1}{%
\def\temp{#1}\ifx\temp\empty%
\else%
\linelabel{line:#1}%
\fi%
\textcolor{red}{#2}}%
}

%% file: supp.tex
\newpage
\appendix
\nolinenumbers

\setcounter{section}{0}
\setcounter{figure}{0}
\setcounter{table}{0}
\setcounter{linenumber}{1} 

\renewcommand{\thefigure}{S\arabic{figure}}
\renewcommand{\thetable}{S\arabic{table}}
\renewcommand{\thesection}{S\arabic{section}}

\setcounter{linenumber}{1} 
\linenumbers

\title{ - Supplementary Material - \\
An evaluation framework for sparse 4D (3D + time) imaging reconstruction via 
bootstrapped cross-validation
}
\author{Yuhe Zhang$^{1,2}$, Zisheng Yao$^1$, Zhe Hu$^1$， Tobias Ritschel$^3$, \\ and Pablo Villanueva-Perez$^1$}
\date{%
    $^1$Synchrotron Radiation Research and NanoLund, Lund University, Box 118, 221 00， Lund, Sweden\\%
    $^2$Medical Radiation Physics, Lund University, Box 117, 221 00, Lund, Sweden\\%
    $^3$University College London, WC1E 6BT London, UK\\[2ex]%
}
\maketitle

\section{Non-interlaced cross-validation analysis for sparse 4D data}
\label{sec:Non-interlaced}
In the main text, cross-validation is performed using interlaced time points to ensure temporal independence between the paired 4D inputs. 
As described in Equation \ref{Eq:interlaced}, correlations are computed between $\hat{\mathbf{y}}_a^i|_{t=2N}$ and $\hat{\mathbf{y}}_b^i|_{t=2N+1}$, where $\hat{\mathbf{y}}_a^i$ and $\hat{\mathbf{y}}_b^i$ denote an example pair of subset reconstructions selected from the 1000 cross-validation pairs.
This interlacing ensures that the two reconstructions being compared are independent in both space (different projection subsets) and time (different temporal samples), which is required for a valid 4D cross-validation.

For completeness, we also report a non-interlaced comparison, in which correlations are computed directly between $\hat{\mathbf{y}}_a^i$ and $\hat{\mathbf{y}}_b^i$ without temporal offset. In this case, the paired inputs are independent only in the spatial domain (3D) but not in time. The results are shown in Figure~\ref{fig:CV_Sparse}.

\mycfigure{CV_Sparse}{
Performance evaluation of 4D reconstructions under sparse spatial sampling without interlaced time points.
Results are shown as a function of the number of projections for six metrics: MSE, DSSIM, FHC, PSNR, NCC, and NMI (error bars indicate standard deviations).
Three comparisons are reported: subset reconstruction $\hat{\mathbf{y}}^i$ versus ground truth \groundTruth (circles, blue solid lines), subset reconstruction $\hat{\mathbf{y}}^i$ versus the full 16-projection reconstruction \fullsetRecon (triangles, yellow dashdotted lines), and cross-validation between independently reconstructed $\hat{\mathbf{y}}^i$ volumes (squares, green dotted lines).
Because no interlacing is used, the paired inputs for cross-validation are only 3D-independent.
The gray dashed line denotes the performance of the full 16-projection reconstruction \fullsetRecon relative to the ground truth \groundTruth.
}

As can be seen, the lack of temporal independence affects the resulting cross-validation behavior, represented by the green dotted curves in Figure~\ref{fig:CV_Sparse}.
Spatially uncorrelated but temporally correlated inputs tend to exhibit artificially high agreement because the two reconstructions share the same underlying physical evolution. 
Consequently, the cross-validation curves in the non-interlaced case are systematically higher than those in the interlaced case shown in Figure~\ref {fig:IT_CV_Sparse} in the main text.
In contrast, enforcing independence in 4D (both space and time) provides a stricter estimator of true 4D reconstruction fidelity.
The discrepancies between the two settings, especially at low projection counts, highlight that valid 4D cross-validation requires independence in both space and time. Relying on 3D-only independence may lead to overly optimistic performance estimates.

\newpage

\clearpage

\section{List of Supplementary Data}
\label{sec:appendix}

\begin{table}[h]
\centering
\caption{Projection angles used for the four-view ultra-sparse sampling across the 16 experiments. Each row lists the four angles (in degrees) assigned to one experiment.}
\label{tab:angles_appendix}
\begin{tabular}{ccccc}
\toprule
\textbf{Exp.} & \textbf{Angle 1} & \textbf{Angle 2} & \textbf{Angle 3} & \textbf{Angle 4} \\
\midrule
1  & 0.00  & 45.00  & 90.00  & 135.00 \\
2  & 11.25 & 56.25  & 101.25 & 146.25 \\
3  & 22.50 & 67.50  & 112.50 & 157.50 \\
4  & 33.75 & 78.75  & 123.75 & 168.75 \\
5  & 45.00 & 90.00  & 135.00 & 0.00 \\
6  & 56.25 & 101.25 & 146.25 & 11.25 \\
7  & 67.50 & 112.50 & 157.50 & 22.50 \\
8  & 78.75 & 123.75 & 168.75 & 33.75 \\
9  & 0.00  & 45.00  & 90.00  & 135.00 \\
10 & 11.25 & 56.25  & 101.25 & 146.25 \\
11 & 22.50 & 67.50  & 112.50 & 157.50 \\
12 & 33.75 & 78.75  & 123.75 & 168.75 \\
13 & 45.00 & 90.00  & 135.00 & 0.00 \\
14 & 56.25 & 101.25 & 146.25 & 11.25 \\
15 & 67.50 & 112.50 & 157.50 & 22.50 \\
16 & 78.75 & 123.75 & 168.75 & 33.75 \\
\bottomrule
\end{tabular}
\end{table}